\documentstyle[emulateapj,psfig,onecolfloat]{article}

\def\etal{{\rm et~al.\ }}
\def\hmpc{\;h^{-1}{\rm Mpc}}

\def\kms{{\rm \;km\;s^{-1}}}

\def\lsun{{\rm L_{\odot}}}

\def\lya{Ly$\alpha$}
\def\lyb{Ly$\beta$}

\def\simlt{\lower.5ex\hbox{$\; \buildrel < \over \sim \;$}}
\def\simgt{\lower.5ex\hbox{$\; \buildrel > \over \sim \;$}}

\begin{document}

\twocolumn[

\title{
On the Search for Quasar Light Echoes
}

\author{
Eli Visbal$^{1,2}$ and Rupert A.C. Croft$^{1}$
}

\begin{abstract}
The UV radiation from a quasar 
leaves a characteristic pattern in the distribution of 
ionized hydrogen throughout the surrounding space.
  This pattern or light echo propagates through
the intergalactic medium at the speed of light, and
can be observed by its imprint on  
the \lya\ forest spectra of background sources. As the echo
persists after the quasar has switched off, it offers  the
possibility of searching for dead quasars, and constraining
their luminosities and lifetimes.  We outline 
a technique to search for and characterize these light 
echoes. To test the method, we create artificial \lya\ forest spectra from
 cosmological simulations at $z=3$, apply light echoes and
 search for them. We show how the simulations can also be used to 
quantify the significance level of any detection.
We find that light echoes from the brightest quasars could
be found in observational data. With absorption line
spectra of 100 redshift $z\sim 3-3.5$
quasars or galaxies in a 1 square degree
area, we expect that $\sim 10$ echoes from quasars 
with B band luminosities $L_{B}=3\times10^{45}$ ergs$^{-1}$ exist that
could 
be found at 95 \% confidence, 
assuming a quasar lifetime of $\sim 10^{7}$ yr. 
Even a null result from such a search would have interesting implications
for our understanding of quasar luminosities and lifetimes.
 
\end{abstract}

]

\footnotetext[1]{
Dept.   of  Physics,   Carnegie   Mellon  University,
Pittsburgh, PA 15213}
\footnotetext[2]{evisbal@andrew.cmu.edu}

\section{Introduction}
The \lya\ forest in quasar spectra (see e.g. Rauch 1998 for a review)
offers a useful means to probe the density and ionization structure 
of the high redshift intergalactic medium. The predictions of cosmological
simulations (Cen \etal 1994; Zhang, Anninos, \& Norman 1995;
Petitjean, M\"ucket, \& Kates 1995; Hernquist \etal 1996; Katz \etal
1996; Wadsley \& Bond 1997; Theuns \etal 1998; Dav\'e et al. 1999)
have been seen to match observational
data well (e.g., Viel \etal 2004).
In current cosmological theories, the \lya\ forest is
produced by the remnant neutral hydrogen in a smoothly fluctuating
density field of largely  photoionized gas. 
The fluctuations
in the \lya\ optical depth
$\tau$ are predicted (see e.g., Weinberg \etal 1997, Croft \etal 1997) 
to be related to the gas density ($\rho$)
at each point (${\bf x}$) and inversely proportional to the
intensity $J$ of the ionizing background radiation field:
\begin{equation}
\tau\propto \rho({\bf x})^{\sim 1.6}/{J({\bf x})}
\label{fgpa}
\end{equation}
 Close to the brightest sources, quasars, the effect
of increased $J$ is strong enough that it has been possible to detect
the decrease in \lya\ forest absorption along the
same quasar sightline, the so-called proximity effect 
(Bajtlik \etal 1988, Scott \etal 2000). 
Quasar radiation is expected to travel not only along the line of sight, but
also transverse to it, and affect the ionization state of the IGM
which can be probed by adjacent sightlines. This effect, known as
the transverse, or foreground proximity effect has not yet been seen 
convincingly in hydrogen \lya\ forest data (e.g., Schirber \etal 2004,
 Croft 2004,
although see Dobrzycki \& Bechtold 1991).  

In general, the presence of discrete sources of photoionizing 
radiation will lead to spatial fluctuations in $J({\bf x})$.
The statistical properties of these have been explored by many authors,
including for example Zuo (1992) and Meiksin \& White (2003).
Strong sources such as quasars will leave particularly
recognizable imprints in the intergalactic radiation field (see Figure
5 of Croft 2004). Their size, shape and strength can be probed
using multiple \lya\ spectrum sightlines, in a generalization
of the transverse proximity effect.
 Adelberger (2004)
has shown how sightlines chosen around observed quasars can be used to
measure the radiative histories of quasars through the transverse
proximity effect. We choose to use the term ``light echo'' to describe
these features because although they are not directly analogous
to supernova light echoes (e.g., Crotts \etal 1989) the light
from quasars is still detectable through them after the quasars themselves 
have become ``quiet''.
 We will explain in \S2 how we can find the light echoes and
so detect quasars and
yield constraints on their properties even when the light from 
them is not directly observable.

There are several examples of extended objects which have been searched
for in cosmological datasets with some type of matched filtering.
These include galaxy clusters at high redshift (Postman \etal 1996).
Circles in the CMB sky, which would be a sign of a universe with a small
topology scale have also been looked for (Cornish \etal 2004). We will 
use a similar idea to find and measure light echoes in \lya\ forest data,
searching for the signature of a deficit of \lya\ absorption with a 
particular geometry.

The lifetimes of quasars are relatively poorly constrained, and
 observationally the evidence points to the range $10^6-10^8$ Yr (see
Martini 2003 for a review). Theoretical predictions of quasar lifetimes
have recently become available from hydrodynamic simulations of 
galaxy formation which include black hole accretion and feedback 
(e.g., Di Matteo \etal 2005, Hopkins \etal 2005). Successfully
measuring the transverse proximity effect or finding a light echo
would represent one way to test these models.

This paper is structured as follows.  
In \S2 quasar light echoes are outlined in more detail. In \S3 
the simulations used 
to produce spectra to which light echoes were artificially applied 
are described.  In \S4 a technique is described to search for 
quasar light echoes in \lya\ spectra.  In \S5 the results of
 searches for light echoes in simulated data are presented. These 
results include the sensitivity of the test to different quasar luminosities
 and to varying the number of and resolution of \lya\ spectra.  
In \S6, we discuss our results, computing the 
chances of finding a light echo in observational data and the volume of data 
that would be required to do so. 

\section{Quasar Light Echoes}

The radiation emitted from a quasar has an effect 
on the ionization state of the gas through which it 
passes.  If a quasar produced high levels of radiation 
for some period of time a signature will be left in 
the surrounding gas long after this period has stopped.  
There will be lower levels of neutral hydrogen in the 
region affected by the propagating radiation from the 
once active quasar, as described by Equation \ref{fgpa}.
 The equilibration time, the time taken for the ionization
state to respond to small changes (factors of a few) in the 
intensity of the ionizing radiation is of the order of $10^{4}$ yr
(Martini 2003).
As this is much shorter than the quasar lifetimes we will be considering,
 the effect of the quasar radiation will effectively propagate through
the intergalactic medium at the speed of light. The width of the
light echo will be equal to the light travel time multiplied by the
length of time the quasar was radiating.

\begin{figure}[t]
\centering

\psfig{file=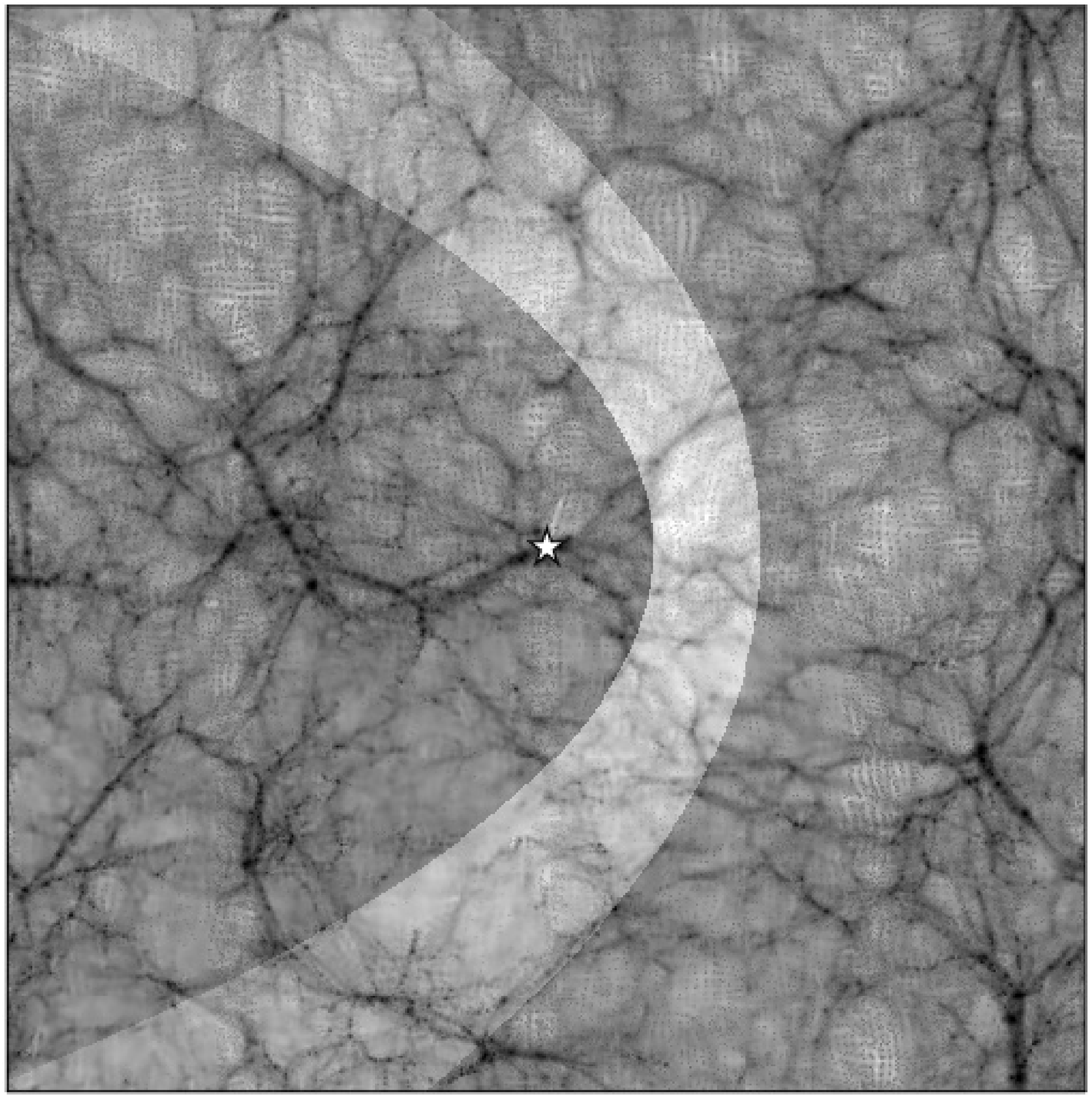,angle=-90.,width=8.5truecm}
\psfig{file=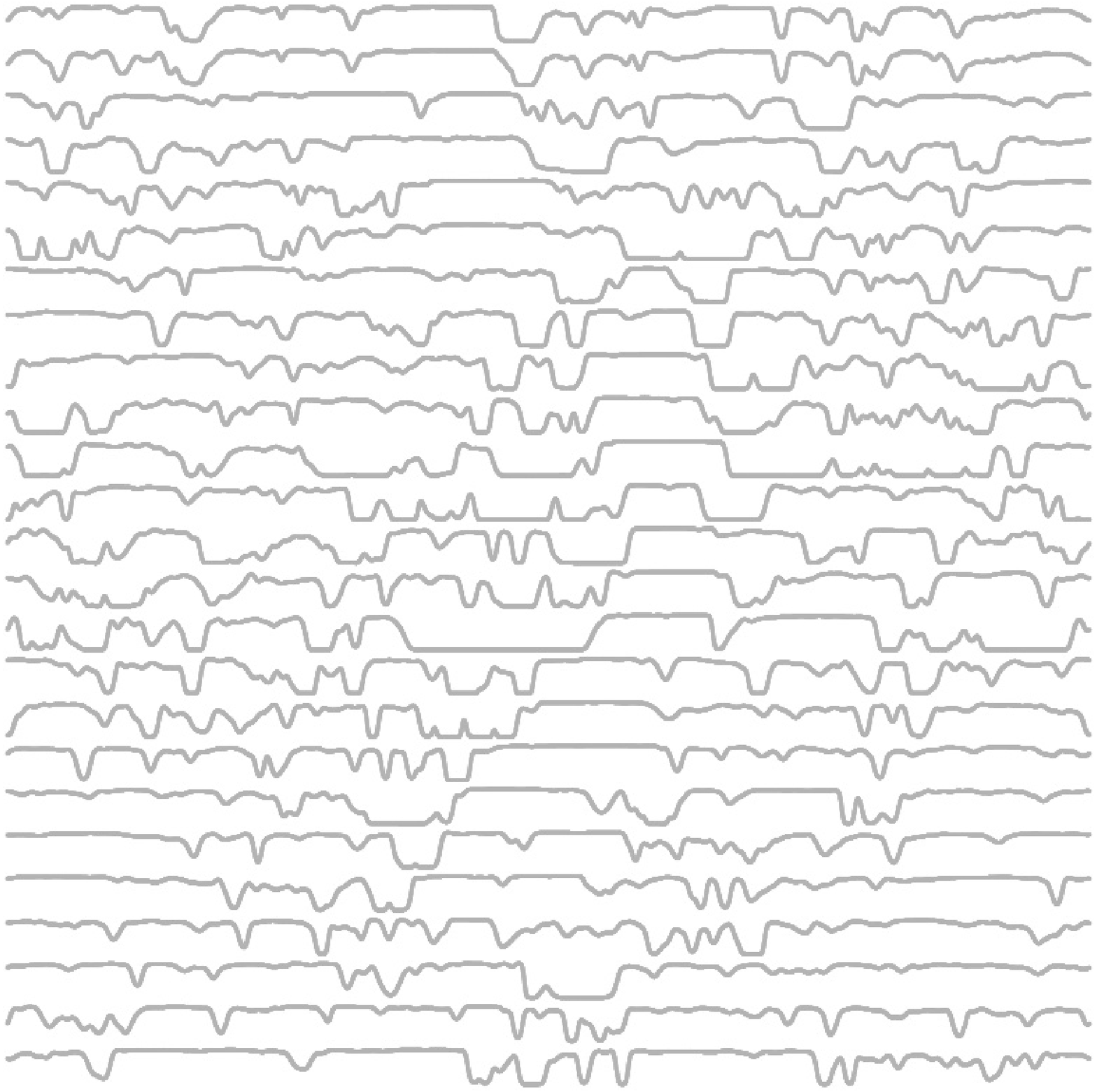,angle=-90.,width=8.5truecm}

\caption[shape]{
\label{spape}
Top panel: The neutral hydrogen density in a $1 \hmpc$ thick slice through a
simulation of a $\Lambda$CDM universe. The plot width is $50$
comoving $\hmpc$. A quasar light echo has been applied and can be seen, the
quasar (position shown by a star symbol) having a lifetime $t_{q}=8.2$ My and
switching off 8.2 My ago.
The observer is observing from
the left side of the plot, the line-of-sight being the $x-$axis
Bottom panel: 25 \lya\ spectra have been made in 
 the plane of the top figure, showing the effect of the
 quasar light echo.  This arrangement of spectra is
a very artificial situation,
 but illustrates how a light echo can be seen.  The high 
flux regions in the spectra correspond to areas of low neutral 
hydrogen density caused by the light echo. }
\end{figure}

  For simplicity, in the present paper we approximate the
quasar lightcurves by a top hat i.e.,  we assume that a
quasar starts to emit a constant level of radiation and
 stops sharply some time $t_{q}$ later.  We will also
assume that this takes place at a redshift where the
expansion timescale is significantly less than $t_{q}$,
so that we can model the quasar radiation using the
inverse square law. An additional simplification we use is
to neglect the attenuation of the quasar due to intergalactic 
absorption. As the attenuation length at $z=3$ is of the order
of $100 \hmpc$ (Haardt \& Madau 1996) this is not a bad approximation.

If we were able to recieve information from all points 
in space at the same time, a light echo would be a spherical shell.
 The outer boundary would correspond to the start of emission
 and the inner one correspond to the end, with a separation
$ct_{q}$ between them.
However, when observing a light echo one would not be able to see every point 
in space at the same time.  The boundaries of the regions containing 
radiation can be described as follows.  Given a Cartesian coordinate
system with the $z-$axis being oriented
directly away from the observer and the origin 
centered on the quasar: 

\begin{equation}
   R_{on} = ct_{on}-z,
\end{equation}
 where $R_{on}$ is the distance the light has traveled since the start of 
radiation emission and $t_{on}$ is the time since the start of this emission
considered at the location of the quasar.

Changing to spherical coordinates one obtains

\begin{equation}
   R_{on} = ct_{on}-R_{on}\cos(\theta) 
\end{equation}
 Thus, the surface corresponding to the 
start of the radiation can be described as

\begin{equation}
 R_{on} = \frac{ct_{on}}{1+\cos(\theta)}
\end{equation}
 
Another surface can be constructed in exactly the same way 
using, $t_{\rm off}$, the time since
 the quasar emission has stopped considered at the 
location of the quasar.  The 
light echo is then the region enclosed by the two 
surfaces $R_{on}$ and $R_{\rm off}$.

There are six parameters which describe the precise shape of 
the quasar light echo: the length of time since the quasar started 
and stopped emitting radiation, the three dimensional coordinates 
of the quasar and the luminosity of the quasar. 

 In the  top panel of Figure \ref{spape}, we show an example of a
light echo from a bright quasar applied to 
a slice from a cosmological simulation (described in \S3).
 The neutral hydrogen density is shown
as shades of gray, and the shape of the light echo can be seen clearly.

In the examples in this paper, we assume that the quasar radiates
its energy isotropically. Unified models of Active Galactic Nuclei 
(e.g., Urry \& Padovani 1995), however predict that the ionizing radiation may
be beamed into a cone with opening angle $\sim 90 \deg$ (a solid
angle of $\sim 1.8$ rad.) This cone would 
restrict the geometry of the light echo. In this case, the signal to noise
ratio of a light echo detection would be reduced by approximately
$\sim \sqrt(1.8/4\pi) \sim 0.4$ unless 
extra parameters were introduced to model the beaming.
We leave investigation of this possibility to future work (see also Croft
2004, Adelberger 2004.)

\section{Simulations}

We use cosmological N-body simulations of a $\Lambda$CDM universe
in order to develop and test our light echo search technique.
The cosmological parameters we assume are consistent
with the first year WMAP results (Spergel \etal 2003) and 
are Hubble constant  $H_{0}=70 \kms {\rm Mpc}^{-1}$,
 $\Omega_M=0.3$, $\Omega_{\Lambda}=0.7$,
amplitude of mass fluctuations,  $\sigma_{8}=0.9$. 
Our simulations are dark matter only, run with the N-body code 
Gadget (Springel, Yoshida \& White 2001) in a periodic cubical 
volume of side length $50 \hmpc$, ($h=H_{0}/100 \kms  {\rm Mpc}^{-1}$) 
with $256^{3}$
particles. We carry out 20 runs with different random phases,
and use output snapshots at redshift $z=3$.

We make \lya\ spectra from the simulations by assigning an SPH-like smoothing
kernel to each dark matter particle to mimic the distribution of gas
in the IGM. We then integrate along sightlines through the kernels in 
the usual manner
(e.g., Hernquist \etal 1996), computing the density in pixels. We 
use the Fluctuating Gunn-Peterson
 Approximation (e.g., Weinberg \etal 1997) to  assign
a \lya\ optical depth to each pixel, and a power-law temperature density
relation $T=T_{0}(\rho/\overline{\rho})^{0.6}$  
 to assign temperature. 
Here $T_{0}=$20000 K (see e.g., Schaye \etal 2000) and $\overline{\rho}$
is the mean density of baryons in the Universe. We note that the
adiabatic index is likely to be lower than $0.6$ at $z=3$ (observations
such as Schaye \etal 2000
suggest an isothermal equation of state.) The large scale 
fluctuations in the \lya\ forest are however insensitive to this
choice (see e.g., Fig. 7 of Croft \etal 1998.)
We convolve the real space
distribution of optical depths with the thermal broadening and 
the line-of-sight
peculiar velocity field to produce spectra in observable units (in redshift
space).

In carrying out this procedure, we first assume a uniform ionizing radiation
field throughout the volume. We normalize the $\tau$ values so that the
value of the mean transmitted
flux $\langle F \rangle=0.696$, that observed by Schaye
\etal (2003). We also make spectra with light echoes. In this case, we
choose the position of the quasar (in this paper, we pick random locations)
and apply
the light echo to the real space optical depths in the
 spectra using Equations 1-4. After the light echo has been applied we
 convolve the spectra with thermal broadening and peculiar velocities.

In the bottom panel of Figure \ref{spape} we show an example of 25 spectra
that correspond to the neutral hydrogen density field in the simulation
slice above it. In this case, the sightlines were taken to all lie in a
plane. This is an obviously artificial situation for illustrative
purposes only, and in the rest of the paper
we assume that the sightlines are randomly distributed.
We can see the outline of the light echo in the  
 \lya\ forest of Figure \ref{spape}, but it is difficult to distinguish
by eye. We
shall see below that by passing a template through the datasets we can 
detect the light echoes and compute their significance level.

Using the spectra in the simulations, we construct fiducial artificial
datasets, all at redshift $z=3$.
 Our fiducial datasets contain 50 spectra each and
we rebin the high resolution pixels to 50 pixels per spectrum, corresponding
to a pixel size of 1.8 Angstroms.
In order to determine the statistical significance of light echo 
detections, as detailed in our method below, it is necessary to 
create 1000 artificial datasets. We do this by computing 50 sets of
spectra (always with the same positions, initially randomly chosen),
for each of the 20 simulations with random phases, but for each set randomly
translating the box, rotating it through a multiple of 90 degrees about the
 3 axes and randomly reflecting it.

\section{Search Technique}

 To search for a quasar light echo in a set of spectra, we create templates  
 by applying a light echo artificially to uniform spectra (all 
pixels initially have transmitted flux $F=\langle F \rangle$ with 
the same spatial configurations.)  The data being searched is then 
compared to the templates.  A value of $\chi^{2}$ is then computed
by comparing pixels in the template with those
in the dataset:
\begin{equation}
\chi_{{\rm pre-norm}}^2 = \displaystyle\sum_{i=1}^{N} 
\displaystyle\sum_{j=1}^{M}\left(\frac{F_{\rm temp}-F_{\rm data}}
{\sigma}\right)^2 
\label{chi2}
\end{equation}
Here $N$ is the number of spectra, $M$ is the 
number of pixels per spectrum, $F_{\rm temp}=e^{-\tau_{\rm temp}}$,
and  $F_{\rm data}=e^{-\tau_{\rm data}}$, are the transmitted
flux in pixels for the template and the data respectively.
The standard deviation of the
pixel values $F_{\rm data}$ in the data is used
to compute $\sigma$. Our estimate of the ``noise'' will therefore
be dominated by the intrinsic density fluctuations in the 
transmitted flux. We do not however compute the full covariance matrix,
but instead will compute the significance of light echo detections
by looking at the probablity of false detections in simulated datasets
with no light echoes (see below).

\begin{figure}[t]
\begin{centering}
\psfig{file=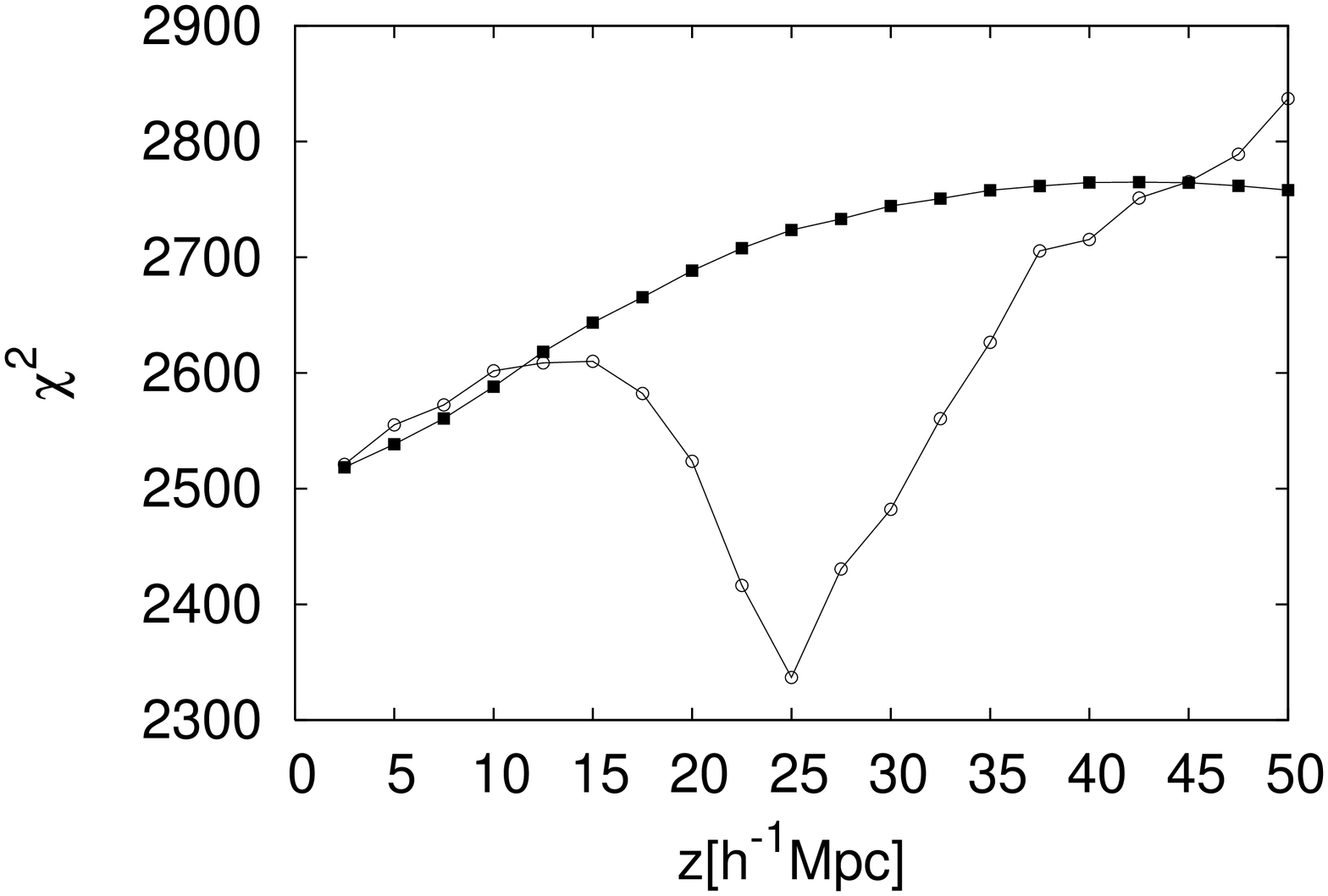,angle=-90.,width=8.5truecm}
\psfig{file=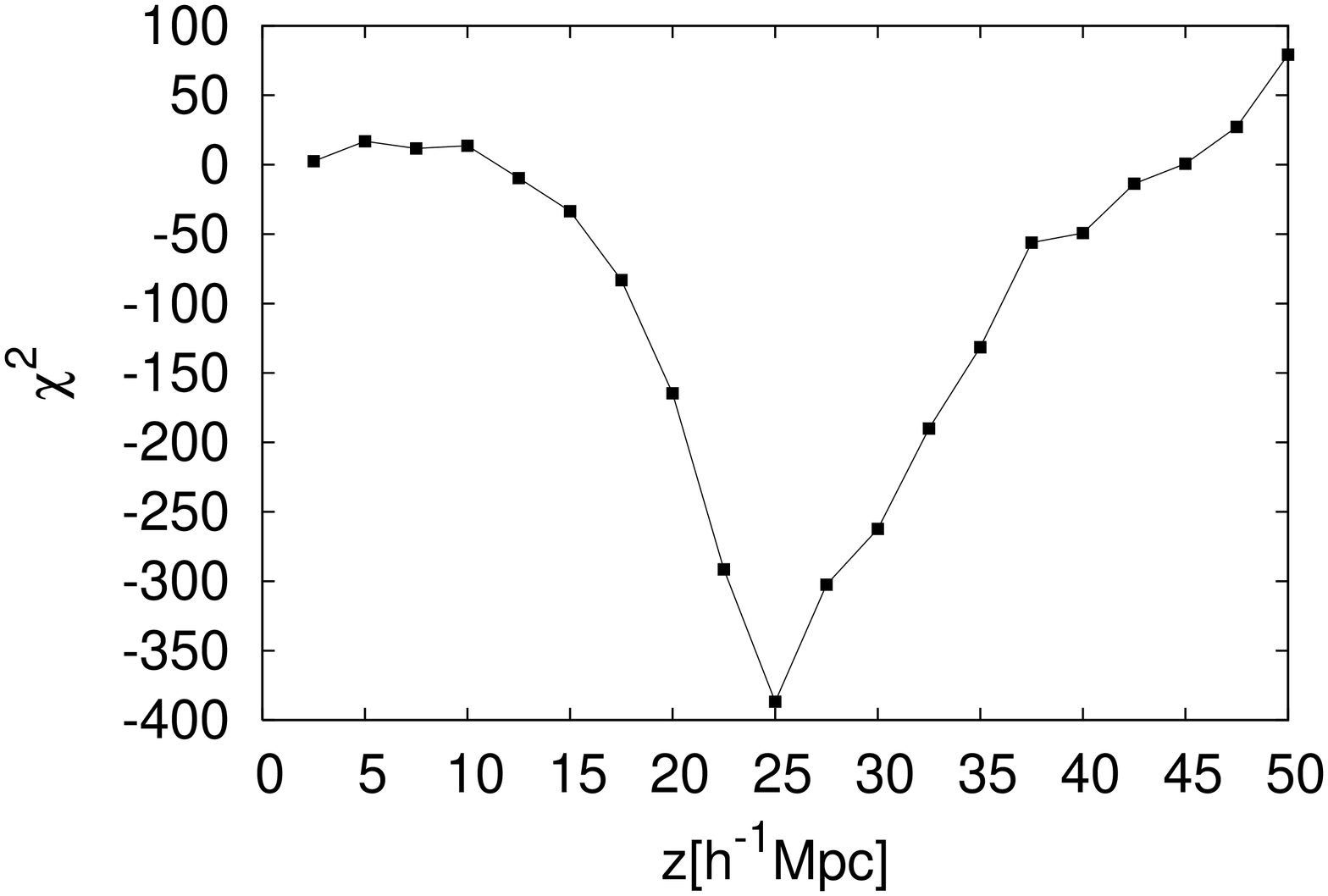,angle=-90.,width=8.5truecm}
\end{centering}
\caption[likely]{
\label{likely}
A search for a light echo in one dimension. 
We plot the  $\chi^2$ that results when
fitting light echo templates with a  varying $z$ coordinate for
the quasar position (distance from the
boundary
 of the simulation box along along the line of sight to the observer).  
We fit the templates to a simulated observational dataset with 
an imposed  light echo (true quasar position is $z=25 \hmpc$).
 The top panel shows the results (as circles)
before subtracting the background 
average $\chi^2$ (this background is
also shown as filled squares) while the bottom panel shows the
renormalized results  (see \S 4 for details).
 }
\end{figure}

When we compute the $\chi^{2}$ value in Equation \ref{chi2}, 
we make sure to compute the standard deviation $\sigma$ for each 
data set separately. Without doing this, we find that the 
significance of detections is degraded by an order of
magnitude or more.

In our technique, a
 grid of values for each parameter describing a light echo is set up
and a template is made for each possible combination of these values.  
If a template has a very low $\chi^2$ it is likely that there is 
a quasar light echo in the data with the same parameters as the template.

The less space an echo occupies in the template (for example
if the echo is on the far edge of the simulation
box closest to the observer), the lower the  $\chi^2$
values tend to be.  To account for this it is necessary to 
renormalize the likelihoods.
To do this, we find the $\chi^2$ values of all the templates fitted
to many mock spectra generated without applied light echoes and 
average them.  This average  $\chi^2$  is then subtracted from the 
pre-normalized  $\chi^2$ values determined as described above giving:
\begin{equation}
\chi^2=\chi_{\rm pre-norm}^2-\chi_{\rm avg}^2
\end{equation}
By doing this we are subtracting the
distribution of $\chi^2$ values which arises purely from the
geometry of the sample, to reveal 
the true light echo signal.  After doing this, we
identify the templates with the lowest  $\chi^2$ values.
Because we have only applied one light echo to the simulated data and 
are searching for that, we associate  the minimum  $\chi^2$ value
with the echo we are searching for. In a real observational dataset,
our search could include the possibility  of multiple minima in the 
$\chi^2$, corresponding to the presence of several light echoes.

After the lowest  $\chi^2$ is found, we determine the 
statistical significance of the detection. We do this by 
applying the same search procedure to our 1000 simulated datasets, but without
having applied light echoes to them.  The statistical 
significance is the probability that an equal or lower $\chi^2$ could 
be caused by statistical fluctuations. These statistical
fluctuations correspond to density fluctuations which by chance mimic
the geometry of a light echo. These could occur anywhere in the
simulation volume.   

In summary the light echo search technique consists of the following steps:\\
1.  Pick a range of values and create a grid of the six parameters 
($x,y,z, t_{\rm off}, t_{\rm on}, L$) describing a light echo.\\  
2.  Create a template for every point on this grid and compare with 
data to calculate $\chi^{2}$ for each.\\
3.  Subtract average $\chi^2$ computed from 
 many simulated spectra without light echoes to obtain normalized $\chi^2$\\
4.  Find $\chi^2$ minimum.\\
5.  Create many simulated spectra with the same coordinates, but without 
a light echo.\\  
6.  Repeat steps 3-5 on each set of simulated spectra and count fraction 
of cases with lower $\chi^2$ to determine statistical significance.

\section{Tests and results}

\begin{figure}[t]
\centering
\psfig{file=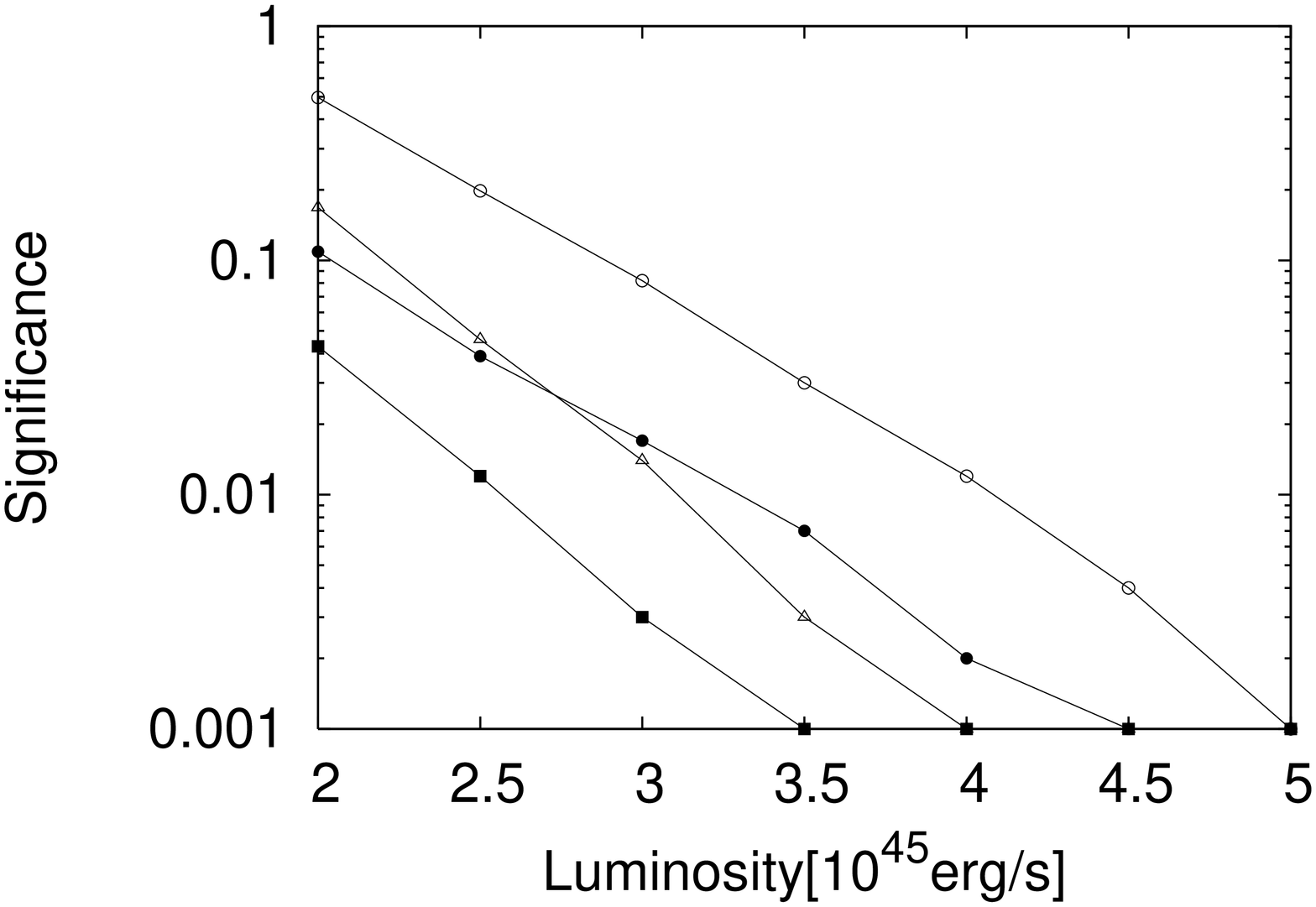,angle=-90.,width=8.5truecm}
\psfig{file=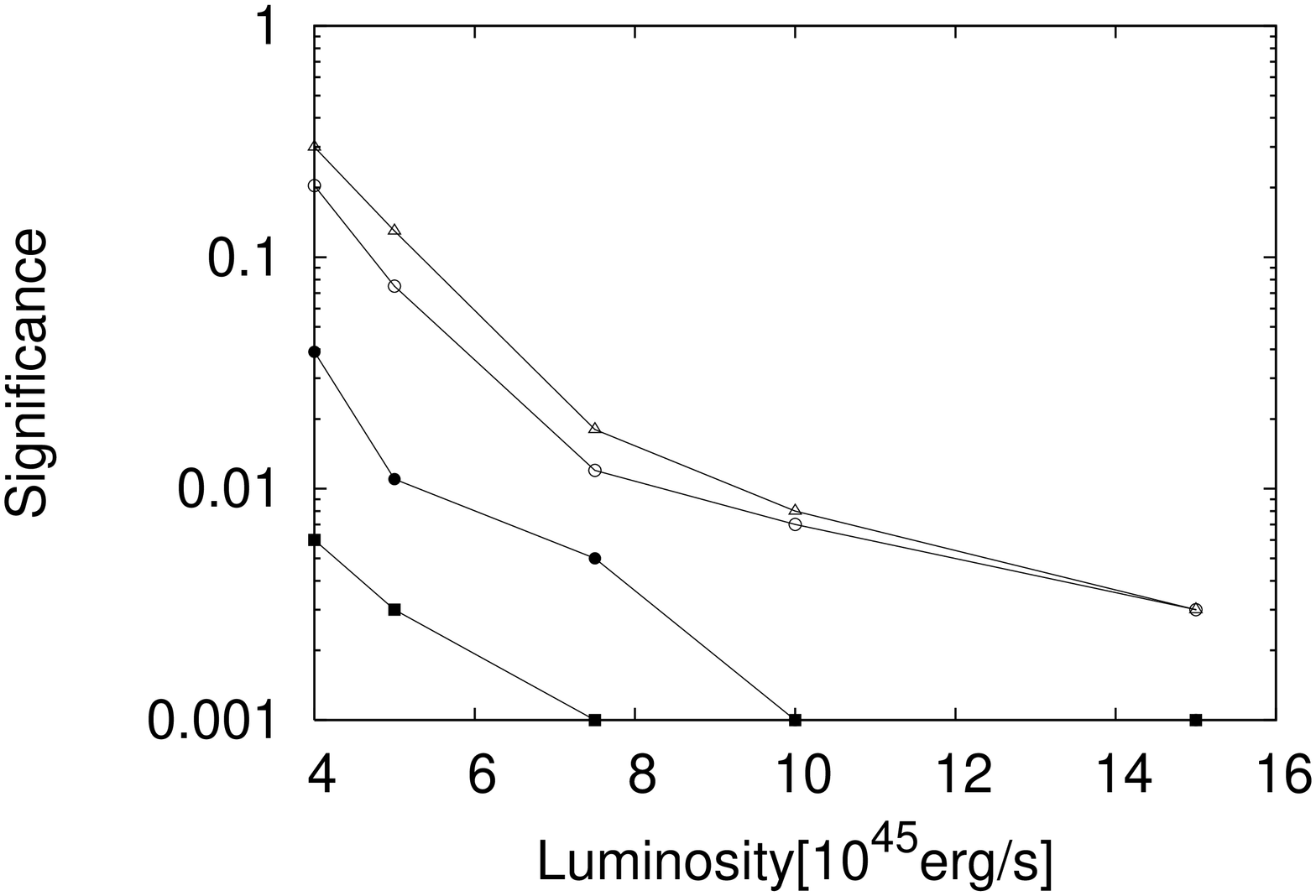,angle=-90.,width=8.5truecm}
\caption[lum]{
\label{lum}
Statistical significances for
detection of light echoes in mock observational data
(see \S5) as a function of quasar luminosity.
The four different curves correpond to four different simulated
datasets with different random seeds.
In the top panel the light echoes being searched for all had a lifetime
$t_{q}=16.3$ Myr and a time since the end of radiation emission of
16.3 Myr (see text: case A.)  Different sets of mock observational 
data were made with
different quasar blue band
luminosities and the statistical significance of detection
is plotted versus that luminosity. In the bottom panel (text: case B)
the light echoes
in the mock data all had lifetime $t_{q}=32.6$ Myr and time since the 
end of emission 65.2 Myr (i.e. the light echo was much farther from 
the position of the quasar than in the top panel).
 }
\end{figure}

We set up a test light echo inside the cubical volume of the simulation and
use our search technique described above to find and parametrize it.
The location of the quasar was chosen to be in the center of the box.  
The duration of time since the beginning and end
of the radiation emission were set to be 32.6
 Myr and 16.3 Myr respectively ($t_{\rm on}$ and $t_{\rm off}$ in Eqns. 2-4). 
The quasar lifetime ($t_{\rm on}-t_{\rm off}$)
 in this case was therefore 16.3 Myr. We refer to this as test case A.

In order to make our testing more widely applicable, an additional,
different test (case B) was also set up, this time with the echo located
much further from the quasar position (i.e. the time since
the quasar switched off is much longer). 
This time the 
 quasar is located at $x ,y$ and $z$ coordinates of 
$(25,25,10) \hmpc$, where the $z-axis$ is oriented along the
line of sight towards the observer. 
The $t_{\rm on}$ and $t_{\rm off}$ times were 97.8 Myr 
and 65.2 Myr respectively, so that the  quasar lifetime is
 32.6 Myr, double that of the previous case.

\begin{figure}
\centering
\psfig{file=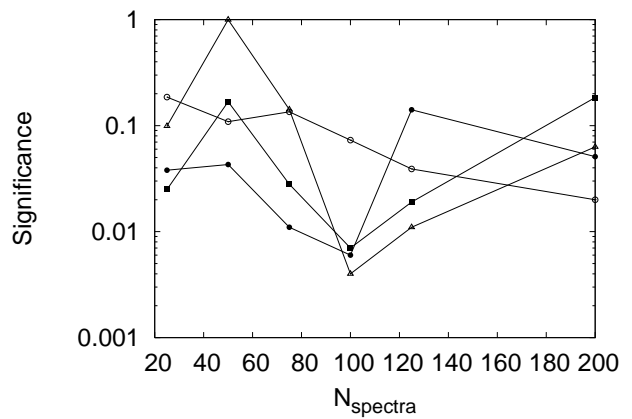,angle=-90.,width=8.5truecm}
\caption[numspec]{
\label{numspec}
Statistical significances of finding a quasar light echo with a 
blue-band luminosity of $2 \times 10^{45}$ erg/s with differing number 
of spectra in the mock observational dataset.  All spectra had 50 pixels
each (1.8 \AA width).
 Four simulations with different random seeds 
 were used to make the 4 different lines.

 }
\end{figure}

For each of these two test cases, we create a simulated dataset,
using a number of spectra which could be achievable with observations (see
\S6.2 for further discussion). As stated in \S3, We pick 50 
spectra passing through the simulation
volume, which subtends an angle on the sky of 39 arcmin at $z=3$. We note
that we only use a fraction  of the information available
in each spectrum as each full \lya\ to \lyb\ region is 7 times
 the length of our simulation box. In \S5.2 below we will investigate
the effect of varying pixel size and number of quasar spectra on 
the ease of detection of light echoes.

We used our 1000 sets of simulated
 spectra to find 
the average background likelihood distribution as well
to calculate the statistical significance of located echoes.
In the tests, we set up a grid of search parameters, with a 
$5 \hmpc$ spacing in the $x,y,z$ coordinates of the quasar,
 and also a 8.2 Myr spacing in   $t_{\rm on}$.
In the present work, we have chosen to not search through the other 
parameters ($t_{\rm off}, L$) at the same time, but instead
assume that they are known. We then vary them individually in later tests
to show that they can be constrained.
We leave it to future work to 
search through the 6 dimensional grid directly. 

In order to deal with the effect of not varying $t_{\rm off}$,
 we assume that a template used to search for light echoes
would be discretized so that $t_{\rm off}- t_{\rm on}=\Delta t,$
where $\Delta t$ is an interval of time equal to the shortest
quasar lifetime to be searched for. In this way only one of the two parameters
$t_{\rm off}, t_{\rm on}$ needs to be searched for directly and if 
the actual quasar lifetime $t_{q}$ is greater than $\Delta t$ then
multiple light echoes will be detected originating in the same place
but at different times. They can be summed together to recover the full echo.

In Figure 2 is an example plot showing
how the $\chi^{2}$ varies as a function 
of coordinate $z$ , distance along the line of sight to the observer,
for all other parameters held fixed at the input values used to
construct the light echo. The panels show the raw  $\chi^{2}$
and the renormalized  $\chi^{2}$, obtained after subtracting the
background average from all the random realizations. 
We can see that the minimum in the $\chi^{2}$ is at the input value, for
this, test case A, $z=25 \hmpc$. 

\subsection{Sensitivity of Technique to Light Echoes of Different Luminosities}

The effect of the light echoes on the \lya\ forest depends on 
the ratio of the quasar radiation intensity to the UV background intensity.
The background in our test cases was set to be 
$5 \times 10^{-22} \frac{erg}{(s)(Hz)(Sr)(cm^2)}$,
consistent with the results of Rauch \etal (1997).

The restframe blue-band luminosity of each quasar was  
calculated by approximating the luminosity per 
frequency interval as being $\propto \nu^{-1}$ 
and integrating over the rest frame Blue-band.
\begin{equation}
L= \int \frac{A}{\nu}d\nu
\end{equation}
  The spectral radiance of ionizing radiation is then compared to that of the 
backround radiation:
\begin{equation}
E_\nu =\frac{A \cdot 91.2 nm}{c \cdot 4\pi(sr)(r/[1+z])^2} 
\end{equation}    
where $A$ is the constant from the luminosity equation above and $r$ is the 
comoving distance from the quasar. 

Quasar light echoes corresponding to quasars of 
different blue-band luminosities were applied to 
identical sets of spectra and then used to determine the 
sensitivity of the search technique.  In order to gauge how the significance
can vary from quasar to quasar we chose 4 different
simulations with different random seeds to make our simulated datasets
and show results for each
of these test cases separately in the plots.

Figure \ref{lum} shows the statistical significance of each 
detected quasar light echo, i.e. the fraction of simulation datasets with no
light echo that gave a lower $\chi^{2}$ somewhere in the parameter
space.

The blue-band luminosity required to detect a light echo, in case A, at better 
than $0.1\%$ statistical significance (i.e. $>99.9\%$ confidence level of
 detection) for the particular shape of echo 
used is between  $3.5-5 \times 10^{45}$ erg/s.
Taking the upper end of this range, this 
 corresponds to an AB magnitude M$_{\rm B} \simeq -25.4$ being needed 
for detection.  For $95 \%$ confidence, we need 
$\sim 3 \times 10^{45}$ erg/s, or M$_{\rm B} \simeq -24.8$.

Additionally we tested case B, the light echo 
 which has a much longer time since the quasar switched off.
In this case, the echo is fainter and 
the B-band luminosity required for $95\%$ confidence
is  $5 \times 10^{45}$ erg/s. For $99.9\%$ confidence the
luminosity in this case was $\sim 10-15 \times 10^{45}$ erg/s.
In two of the simulated datasets
 for case B it was not possible to find the echo
at significance greater $99.7\%$, regardless of luminosity.

In case B there were a number of significant false minima found.  
These generally were accompanied by the correct minima which usually had a 
comparable significance level.  These false minima occured because it is 
possible to have very similar echoes produced by altering the $z$ position and 
the start time of emission simultanously.  Decreasing the time
since radiation  emission started while increasing the $z$ 
position leaves a similarly shaped echo.  At a 
luminosity where the echoes were significant to $0.1\%$, false minima like 
this were lower than the correct minima roughly $25\%$ of the time.
We expect that such a degeneracy between $z$ position 
and $t_{\rm on}$ could be 
recognized in observational data and the multiple minima combined into one. 

\subsection{Sensitivity to Number and Resolution of Spectra}

The effects of a differing number of spectra and resolution of spectra on the 
sensitivity of the technique described above was investigated.   
First, the number of (randomly distributed)
spectra was varied from 25 to 200  while holding 
the spectrum pixelization fixed 
at 50 pixels per spectrum. A light echo with blue band luminosity 
$2 \times 10^{45}$ erg/s was used, located within the box as in test case A.
The statistical significance as a function of number of spectra is shown
in Figure \ref{numspec}. Again we show results
for 4 different simulations. 
Although the results are noisy, it can be
seen that the significance does
appear to improve as the number of spectra is increased. 
Having up to 100 spectra makes a useful difference, and then beyond this
there is no noticeable improvement. This number of spectra corresponds
to density of quasars of $\sim  250$ per square degree.

\begin{figure}[t]
\centering
\psfig{file=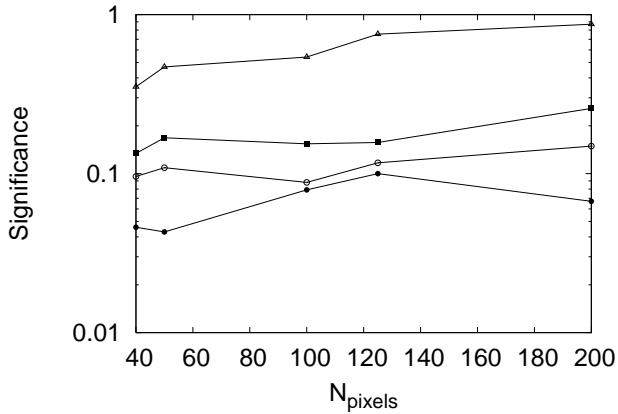,angle=-90.,width=8.5truecm}
\caption[numpix]{
\label{numpix}
Statistical significances of finding a quasar light echo with a 
blue-band luminosity of $2 \times 10^{45}$ erg/s with differing number 
pixels per spectrum in the mock observational dataset.
The number of spectra was held at fixed at 50.
 Four simulations with different random seeds
 were used to make the 4 different lines.
 }
\end{figure}

Next the number of pixels (and hence the resolution) 
was varied, while holding the
number of spectra fixed at 50. The statistical significances for varying 
the numbers of pixels in each spectrum
 can be seen in Figure \ref{numpix}. 
The range shown in the plot
corresponds to $0.45 - 2.25$ \AA\  per pixel.  
We again show results for four different simulations.
 The results seem to 
be rather noisy, with the higher resolution spectra
(more pixels) actually having a 
slightly worse significance level for detection
that lower resolution ones. Although this is rather hard to understand,
 it seems reasonable to infer from this that improving the
spectral resolution below $\sim 2$ \AA will not make data more useful.
 This is probably because the fluctuations in the density field
which play the part of noise in our search for light echoes  have a 
longer correlation length than this so that there is no gain in
increasing spectral resolution. This might not be the case for light 
echoes with very short duration in time (e.g. $< 1$ Myr) which we have
not tested.

\subsection{Remaining Parameters}

In the preceding sections, the parameter search was limited to the
$(x,y,z)$ coordinates of the quasar and one of the
times governing the length of time since the quasar switched off.
To show how the additional
parameters  can be constrained, we search through them independently 
in mock observational data while holding the other five fixed at their
true  values.

\begin{figure}[t]
\centering
\psfig{file=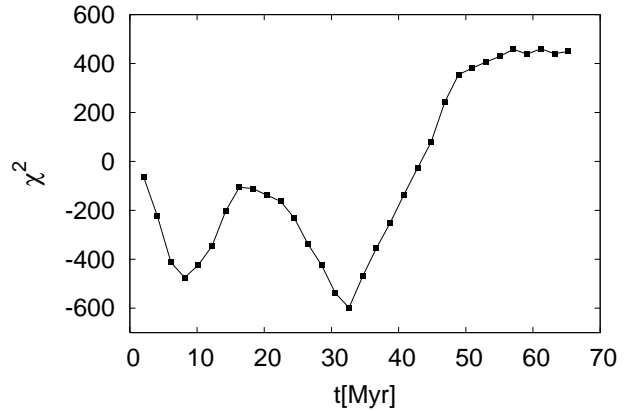,angle=-90.,width=8.5truecm}
\caption[time]{
\label{time}
A 1 dimensional search for the start time of a light echo ($t_{on}$
in Equations 2-4) in a mock observational dataset.
 The other parameters which govern the echo
were set to their true values.
 }
\end{figure}

In Figure \ref{time} we show for light echo test case A how the  $\chi^2$
varies as a function of $t_{\rm on}$. The well-defined minimum is found at the
bin closest to the correct value (32.6 Myr). The other parameter associated 
with the time, $t_{\rm off}$ can also be found in the same way. We
note that our test case quasars had ``top hat'' light curves, but that 
in a more realistic observational case there might be a smoother 
variation with time of the quasar luminosity. With good enough
data this would show up 
as several detected light echoes next to each other
in the data, but with differing luminosities.

\begin{figure}[t]
\centering
\psfig{file=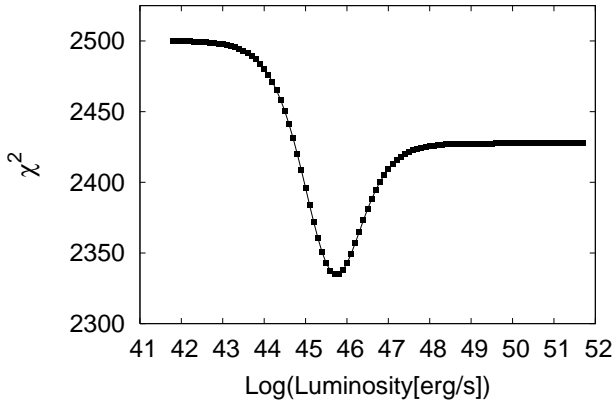,angle=-90.,width=8.5truecm}
\caption[lumsearch]{
\label{lumsearch}
A likelihood search of luminosity in mock observational
data holding other
 parameters fixed to their true values.
 The true luminosity is $5 \times 10^{46} {\rm erg} {\rm s}^{-1}$ 
in the blue-band.  
 }
\end{figure}

In Figure \ref{lumsearch} we vary the luminosity parameter in our search
and show the  $\chi^2$ values. We find that the  $\chi^2$
is approximately constant for a wide range of luminosities
well below the correct one ($5\times 10^{46}$ erg/s, 
with a change starting at about 2 orders of magnitude below it.)
The minimum is found at a value roughly an order of magnitude 
smaller than the actual value, indicating that it is possible that
 the
luminosity of the quasar is more difficult to find accurately than 
the position. The luminosity in this case was very high, though and so
it is possible that the light echo saturated, making it more
difficult to find the luminosity.
A more detailed study of this awaits future work. There is then a
rise in  $\chi^2$ to a plateau. This plateau has a lower $\chi^2$
value that the case of no light echo, indicating that in this case
a template with regions completely devoid of absorption is a worse fit.

\section{Summary and Discussion}

\subsection{Summary}

In this paper we have presented a technique for searching for quasar
light echoes in \lya\ forest spectra. Our conclusions can be
summarized as follows:\\

\noindent(1) We find that our technique which involves
passing a template through a 
realistic simulated
dataset yields a minimum $\chi^{2}$ value for
the appropriate light echo parameters.\\

\noindent(2) The statistical significance of a detection
is strongly dependent on the luminosity of the
quasar and the time since it shut off.
Quasars with B-band luminosity $> 3\times10^{45}$ erg/s
are necessary to make detections at $2 \sigma$
significance or greater when the quasar switched off $\sim 30$ Myr
previously or less.\\

\noindent(3) The number of spectra in a data sample
is also a relatively important factor in the significance
of detection, but their  resolution is not. An
optimal dataset should be densely sampled
with many sightlines per square degree but the signal
to noise or resolution of spectra can be low.

\subsection{Discussion}

\subsubsection{Observational requirements}

We have only simulated the search for quasars light echoes for
quasars with a lifetime of $\sim 10^{7}$ years. We leave it up to future work
to find whether longer lived quasars can be detected more easily
(we suspect that they can, as their signature in the \lya\ forest
will extend much further).
We have seen that a restframe B-band luminosity of
$3\times10^{45}$ erg s$^{-1}$ is necessary to detect light echoes with 
this lifetime at the $2 \sigma$ level, with the significance
increasing if the quasar switched off more recently than 30 Myr previously.
Using observational data for the quasar luminosity function we can compute
for a given lifetime how many quasar light echoes might lie within a
 particular  dataset.

For example, Hopkins \etal (2007) find from a compilation of 
recent quasar luminosity function data that the space density of 
quasars at redshift
$z=3$ with this restframe $B$ band luminosity (corresponding to an
AB magnitude $\sim -24.8$) is $8.1 \times 10^{-7} {\rm Mpc}^{-3}$ (we
use h=0.7 in this calculation).
 We have estimated this
value using the software made available by Hopkins \etal, which includes
data from such samples as Wolf \etal (2003) and  Richards \etal (2006).
If we imagine a survey with a sky area of 2 square degrees, such as the
COSMOS survey (Scoville \etal 2006), then the survey volume 
between $z=2.5$ and $z=3.5$ is $2.3\times10^{7} {\rm Mpc}^{3}$. If we ignore
space density evolution over this redshift range, then we would expect there
to be $\sim 20$ quasars in the volume which reach our $2 \sigma$
detection threshold. However because we are not just able to detect
quasars which are on, we actually expect 
$32.6/16.3=2$ times more
echoes in the survey volume, assuming that quasars have a lifetime
of 16.3 Myr as in our case A test. A shorter lifetime would increase the
ratio of echoes to currently active quasars, although the echoes might
be more difficult to detect, something which should be explored in 
future work. 

For these 40 light echoes in the COSMOS volume to be detectable,
we would need to have a space density of observed 
quasar sightlines comparable to
our simulation tests. Our simulation volume subtends
an angle of 0.4 square degrees and we have used 50 sightlines.
This means that in the 2 square degree survey we
would need absorption line spectra for at least  
250 quasars with redshifts $z\sim 3-3.5$. This is a large number (e.g.,
Prescott \etal 2006 have taken spectra of 95 confirmed quasars in the
COSMOS field, but none with redshifts $z>2.3$).
Of course a survey with a smaller sky area and number of sightlines
 could be chosen, for 
example one with the same footprint as our simulation volume and in which 
$\sim 10$ detectable echoes should be present. 

We note that the quasar lifetime will also govern the angular density of 
sightlines needed to make a detection. Echoes for very short lifetimes would
be better searched for using absorption line spectra of sources with a higher
space density than quasars. For example, Adelberger (2004) has suggested
that the spectra of Lyman break galaxies could be used for the
task of constraining the transverse proximity effect around quasars. 
This technique could also be usefully used for quasar light echoes.
Datasets such as that of Shapley \etal 2006 (deep spectroscopic observations
of star forming galaxies) might be suitable.

At lower redshifts, there will be more quasars available, but the mean 
flux in spectra will be lower, and this will make the echoes more
difficult to detect. In order to gauge this, we have tried carrying out
our search for light echo test case A, but after increasing the value
of $\left<F\right>$ to 0.754, the observed value at $z=2.5$ (from
Press \etal 1993). We find that the
significance of the detection becomes slightly worse (by $35 \%$) on 
average. This is a small effect, that would be more than countered by the 
much larger number of quasar spectra available at this redshift. We note that 
in this test, we did not evolve the simulation to $z=2.5$, so that the 
effect of increased density fluctuations was not modelled. We expect this to
be a relatively small effect also.

\subsubsection{Theoretical expectations}

Additional to assuming a quasar lifetime of 16.3 or 32.6 Myr in our tests, 
we also assumed that the shape of the quasar lightcurve is a simple
step function in time. Our search templates were built with this assumption,
 but it is likely that the actual quasar life history is more complex.
For example, quasars are likely to be triggered by mergers between 
gas rich galaxies and intense outbursts of quasar activity will
occur on each pass as galaxies gradually merge. The merger between 
two large galaxies will take of the order of 1000 Myr, but the 
individual outbursts of quasar radiation which would form individual
echoes will be of much shorter duration. This can be seen in the 
hydrodynamical simulations of galaxy mergers including blackholes
by e.g., Di Matteo \etal (2005), Springel \etal (2005), Hopkins \etal (2005).
In Figure 16 of Springel \etal (2005), it can be seen that during the 
a major merger of two large spiral galaxies
there is a period during which the black hole accretion rate 
rises more than two orders of magnitude above the
background over a period $\sim 100$ Myr, an event which would produce a
strong light echo.

A more sophisticated way of predicting quasar lifetimes from 
these theoretical models involves including the effect of obscuration
during the merger on the lifetime, which for example can shorten the
lifetime measured from the B-Band luminosity  considerably from 
that estimated from the raw black hole accretion rate. This analysis has
been carried out by Hopkins \etal (2005). In that work,
 the lifetime was defined to be the total time spent by a quasar 
above a particular luminosity. This is slightly different to 
the length in time of individual bursts, although in the case
of the brightest luminosities these all generally occur during a single
burst per merger event. Hopkins \etal find that the simulated quasars 
have lifetimes of $\sim 10$ Myr for $L_{B} \sim 10^{11} \lsun$,
with a strong dependence on luminosity (e.g., a lifetime of $\sim 10$
times longer when the luminosity is $\sim10$ times less.)
The limiting luminosity we have found in our tests 
(case A) for $2 \sigma$
detection of light echoes corresponds to 
$1.6\times10^{12} \lsun$ in the B-band, for which the
 predicted quasar lifetimes should be even shorter than $10$ Myr.

Our search technique for light echoes is closely related to the work of 
Adelberger (2004) who described a method for using nearby
sightlines to probe the radiative histories of quasars that are
seen directly in observations. We have effectively extended this 
work to look at the radiative histories of dead quasars by searching
for their light echo signatures.

\subsubsection{Other issues}

In the tests in this paper we have assumed that the 
underlying cosmology is known. Any variation in the geometry of
the Universe caused by different cosmological parameters will 
manifest themselves in a distortion of the light echo shape.
This means that the light echoes could be used in a variant 
of the Alcock-Paczynski (1979) test to measure cosmological
geometry. The anisotropic shapes of 
 reionized bubbles around bright sources at 
higher redshifts has been investigated by Yu (2005). We leave further
research on cosmological constraints that could come from light echoes
to future work. For now we note that because the light echos are 
potentially detectable at great distances from the quasar source
(we have seen here several 10s of $\hmpc$), the effect of 
coherent redshift distortions might be relatively weak and so not 
interfere as much with the measurement as for other methods (e.g.,
 Ballinger \etal 1996).

In addition to quasar radiation, another physical process which could
leave ``gaps'' in the \lya\ forest absorption is the presence of
strong galactic winds. Signs of voids in the neutral hydrogen
distribution around starburst galaxies at $z=3$ have been found 
by Adelberger \etal (2003) amongst others. These regions could in principle
masquerade as small (recently formed) light echos, as the scales
involved are of the order of $1 \hmpc$. Larger scale features caused
by galactic winds are less likely to be mistaken for light echos because
their distorted shape will not be the same, due to the slower than light
speed propagation of winds. If they are present, however, they may 
influence
the significance estimates derived for echos from simulations which 
don't include winds. So far the observational searches for
large scale wind features in the forest have found no
evidence of their existence (Shang, Crotts \& Haiman 2007.)

The light echoes that we are searching for are large-scale features,
and it is a concern that using our limited simulation volume ($50 \hmpc$)
we will underestimate the incidence of large voids, fluctuations
which could mimic a light echo. At this redshift ($z=3$), the
modes on the order of the box size are still linear, so that their
amplitude should still be represented faithfully (unlike the case of
$z=0$ with the same size box, for example.) Nevertheless, the effect of
missing modes on larger scales than the box size should be computed. This would
best be done by carrying out a convergence study with larger boxes,
and will be necessary before making estimates of statistical significances
from observational data.

One thing that we did not test in this paper was whether the
effect of redshift distortions on the light echo template could be
modelled. There were none in the present work, because the quasars
were centered on random locations in the simulation.
In the future it might be possible to make a redshift-distorted 
template, which might slightly improve the significance of 
detections.

In future work, it would also be useful to search through all 6 parameters
which characterize a light echo, in order to make sure that the variations
in the luminosity, which was not searched over here do not affect the
efficacy of the search technique. Also, in the present paper we have 
used the chi-square measured between a template and simulated observations
to signal the presence of a light echo, but have calibrated the 
significance level of a given chi squared using simulations. In the future
it may be possible to calibrate the chi-square versus the
significance level in order to compute the latter more efficiently, 
without needing so many simulations. Testing the method
on mock observations containing several light echoes (e.g., Figure 5(c)
of Croft 2004) would be also be useful step,
as well as trying it out at a slightly
lower redshift ($z=2.5$) where there are more observational data samples.

The question of whether quasar light echoes exist is of course closely
linked to the presence of a transverse proximity effect, quasar radiation
affecting other sightlines. If quasar radiation is emitted very
anisotropically, then this would severely restrict the angular extent
of light echoes. Some hints of this have been seen in the 
anistropic  distribution of optically thick observers by Hennawi \etal (2007).
This would be rather puzzling in the context of unified models of AGN, so that
a search for light echos will have the
potential to reveal much about quasar emission and the absorbing 
gas close to quasars. On the other hand, if the emission is not
so anistropic, light echoes should be detectable. Their unusual nature
may allow their use in interesting tests not only of quasar lifetimes
and radiation output but also cosmic geometry.

\bigskip
\acknowledgments
We thank Kurt Adelberger and Alice Shapley for useful discussions
and Rashid Sunyaev for suggesting that the term light echo
could be used in this context. This work
was supported by NASA Astrophysics Theory grant
NNG 06-GH88G.

\end{document}